\documentclass[aps,prl,twocolumn,showpacs,showkeys,nofootinbib,superscriptaddress,floatfix,10pt]{revtex4-2}

\usepackage[utf8]{inputenc}
\usepackage{graphicx}
\usepackage{amsfonts}
\usepackage{amssymb}
\usepackage{amsmath}
\usepackage[mathscr]{eucal}
\usepackage{setspace}
\usepackage{booktabs}
\usepackage[shellescape,latex]{gmp}
\usepackage{multirow}
\usepackage{placeins}
\usepackage{microtype}
\usepackage[dvipsnames]{xcolor}

\newcommand{\beq}{\begin{eqnarray}}
\newcommand{\eeq}{\end{eqnarray}}
\newcommand{\beqnn}{\begin{eqnarray*}}
\newcommand{\eeqnn}{\end{eqnarray*}}

\newcommand{\OV}{\mathrm{ov}}                 
\newcommand{\stag}{\mathrm{stag}}             
\def\ren{{\scriptscriptstyle{\text{\rm R}}}}        
\def\fir{{\scriptscriptstyle{\text{\rm IR}}}}         
\def\fuv{{\scriptscriptstyle{\text{\rm UV}}}}      

\graphicspath{{figs/}}

\usepackage[normalem]{ulem}

\begin{document}

\makeatletter
\g@addto@macro\bfseries{\boldmath}
\makeatother

\title{Dirac Spectral Density in \ensuremath{N_f \!=\! 2+1} QCD at \ensuremath{T \!=\! 230} MeV}

\author{Andrei\ Alexandru}
\email{aalexan@gwu.edu}
\affiliation{The George Washington University, Washington, DC 20052, USA}

\author{Claudio Bonanno}
\email{claudio.bonanno@csic.es}
\affiliation{Instituto de F\'isica Te\'orica UAM-CSIC, c/ Nicol\'as Cabrera 13-15, Universidad Aut\'onoma de Madrid, Cantoblanco, E-28049 Madrid, Spain}

\author{Massimo D'Elia}
\email{massimo.delia@unipi.it}
\affiliation{Universit\`{a} di Pisa and INFN Sezione di Pisa, I-56127 Pisa, Italy }

\author{Ivan Horv\'ath}
\email{ihorv2@g.uky.edu\\}
\affiliation{University of Kentucky, Lexington, KY 40506, USA}
\affiliation{Nuclear Physics Institute CAS, 25068 \v{R}e\v{z} (Prague), Czech Republic}

\date{Sep 23, 2024}

\begin{abstract}

We compute the renormalized Dirac spectral density in $N_f \!=\! 2+1$ QCD 
at physical quark masses, temperature $T \!=\!230\,$ MeV and system size 
$L_s \!=\! 3.4\,$fm. To that end, we perform a point-wise continuum limit 
of the staggered density in lattice QCD with staggered quarks. We find, for 
the first time, that a clear infrared structure (IR peak) emerges in the density
of Dirac operator describing dynamical quarks. 
We also provide numerical evidence that a component of this peak, which becomes 
dominant in the thermodynamic limit, is due to a non-trivial accumulation 
of near-zero modes. Features of this structure are consistent with those previously 
attributed to the recently-proposed IR phase of thermal QCD. Our results 
(i) provide the only complete first-principles evidence that these IR features exist 
and are physical; 
(ii) improve the upper bound for IR-phase transition temperature 
$T_{\mathrm{IR}}$ so that the new window is $200 < T_{\mathrm{IR}} < 230\,$MeV; 
(iii) are consistent with non-restoration of anomalous U$_{\mathrm A}$(1) symmetry 
(chiral limit) below $T \!=\! 230\,$MeV.

\medskip

\end{abstract}

\maketitle

\noindent
{\bf 1.~Introduction.} 
Dirac spectral density $\rho(\lambda)$ became a useful object in QCD
via realization~\cite{Banks:1979yr} that its strictly infrared 
($\lambda \!\to\! 0$) behavior can be related, in the limit of 
massless quarks, to spontaneous chiral symmetry breaking. However, 
it became clear in recent years, that its physics reach is not only 
in strictly infrared, and that its useful information goes well beyond 
the chiral condensate.

A particularly interesting topic concerns the behavior of $\rho(\lambda)$ 
at high temperatures, namely above the chiral crossover temperature $T_c$.
It was naively expected that, in the chiral limit, $\rho(\lambda)$ vanishes 
near $\lambda \!=\! 0$ due to the chiral symmetry restoration, and that 
the same occurs at real-world light quark masses. However, this conventional 
point of view was challenged by Refs.~\cite{Alexandru:2015fxa, Alexandru:2019gdm} 
and~\cite{Dick:2015twa}. 
Indeed, there key steps were taken concerning the reality of a non-trivial 
accumulation of infrared (IR) Dirac eigenmodes (IR peak), and concerning 
its connection to physics. This crystallized into a new general notion 
of the so-called IR phase~\cite{Alexandru:2015fxa, Alexandru:2019gdm}, 
and a new angle~\cite{Dick:2015twa} on the old problem of thermal 
U$_\mathrm{A}$(1) symmetry restoration in the chiral case~\cite{Pisarski_1984}.

Given the high stakes involved (QCD phase diagram and properties of Quark-Gluon 
Plasma) and the non-perturbative nature of the problem, it is important to collect 
definite evidence about the IR peak via first principles Lattice QCD simulations. 
This is in fact crucial since, at present, its basic IR features are debated at 
the foundational level. The key aspect on the table is whether the relevant IR 
structure exists in the lattice Dirac operator describing physical quarks of
the theory. Indeed, previous evidences favoring a non-trivial IR structure, 
such as~\cite{Alexandru:2014paa, Alexandru:2015fxa, Alexandru:2019gdm, 
Alexandru:2021pap, Alexandru:2021xoi, Meng:2023nxf},  
\cite{Dick:2015twa, Kaczmarek:2021ser}, 
are based on the behavior of the overlap operator~\cite{Neuberger:1997fp}
in lattice formulations where quarks are described by staggered or Wilson 
discretization. In fact, doubts have been expressed about the mere existence 
of the IR peak in the spectral density of dynamical quarks~\cite{Aoki:2020noz}. 
It was also suggested~\cite{Ding:2020xlj} that a dynamical-quark peak may 
exist, but be only realistically observable in certain derivatives of $\rho(\lambda)$. 
Finally, in Ref.~\cite{Kaczmarek:2023bxb} limited preliminary evidence for the peak 
using staggered quarks was pointed out, but no systematic study of its robustness 
and of its continuum scaling were performed, as the paper focused on the study 
of the continuum behavior of the bulk density.

Here we resolve this issue. Indeed, we will show that a cleanly-separated 
dynamical-quark IR peak exists sufficiently close to the continuum limit of 
$N_f \!=\! 2+1$ lattice QCD with staggered quarks at the physical point, and 
temperature $T \!=\! 230\,$MeV. The computed $\rho(\lambda)$ is consistent 
with all spectral features seen in overlap-based~studies. 

Apart from having a close connection to many other previous results
(see e.g.~\cite{Kovacs:2010wx, Giordano:2013taa, Horvath:2018aap, 
Kovacs:2017uiz, Rohrhofer:2019qwq, Cardinali:2021mfh, Horvath:2022ewv, 
Kehr:2023wrs, Alexandru:2023xho, Azcoiti:2023xvu, Kovacs:2023vzi, Giordano:2024jnc}), 
our findings have important physical consequences.  In case of 
the U$_\mathrm{A}$(1) problem they establish, by virtue of a direct first-principles 
calculation, the necessary condition for non-restoration of the anomalous 
symmetry in the chiral limit. Indeed, U$_\mathrm{A}$(1) restoration 
is inherently a quark issue and a non-dynamical evidence is incomplete. 
On the other hand, glue is very important for the predicted IR phase. Here the 
relevant setting is that of real-world light-quark masses where our calculation 
is performed, and where the phase can eventually be experimentally verified. 
Placing our results in that context requires a brief elaboration.

\noindent
{\bf 2.~IR Phase.} 
Rather than focusing just on strictly infrared, the precursor~\cite{Alexandru:2015fxa} 
of IR phase~\cite{Alexandru:2019gdm, Alexandru:2021pap, Alexandru:2021xoi} 
views $\rho(\lambda)$ in its entirety as a distribution of degrees of freedom 
across scales. This work provided evidence that SU(3) gauge theories with 
fundamental quarks involve a parameter region, including high temperature 
QCD, where the abundance of IR degrees of freedom is greatly enhanced. 
This was interpreted as a release of these IR degrees of freedom from 
confinement~\cite{Alexandru:2015fxa}. In this way, the peculiar ``quenching 
artifact" found in pure-glue QCD~\cite{Edwards:1999zm} in the early days 
of overlap operator, became a tool to detect a new phase with (partial) deconfinement 
in a large and physically relevant theory space. In Ref.~\cite{Alexandru:2019gdm} 
this ``anomalous phase" acquired a precise definition and a concrete 
physical meaning. Indeed, a strong negative near-pure power in the overlap IR 
peak ($\rho(\lambda) \!\propto\! \lambda^p$ for $\lambda \!\to\! 0$ 
with $p \!\approx\!-\! 1$) was found in thermal QCD both without quarks and 
with staggered quarks. This led to the definition of IR phase as a regime 
with inverse power-law accumulation of IR modes:
$p \!<\! 0$. The ensuing phase classification reflects 
the thermal {\it state of glue} as follows~\cite{Alexandru:2019gdm}: 
(i) $p \!=\!0$ indicates an \uline{IR-broken} state where IR glue is not decoupled
from the rest, and shows no signs of scale invariance. This is a regular confined 
phase.
(ii) $p \!<\! 0$ indicates an \uline{IR-symmetric} state where IR glue 
decouples, separates out and shows elements of IR scale invariance. This is 
the anomalous (partially deconfined) phase of Ref.~\cite{Alexandru:2015fxa}.
(iii) $p \!>\!0$ features an \uline{IR-trivial} state, where IR glue is again 
inseparable, but the abundance of IR degrees of freedom is power-law 
suppressed. Evidence was given~\cite{Alexandru:2019gdm} that in QCD there 
is temperature $T_\fir$ such that the regime $T \!<\! T_\fir$ is IR-broken, 
$T_\fir \!<\! T \!<\! T_\fuv$ is IR-symmetric (IR phase), and 
$T \!>\! T_\fuv$ is IR-trivial. It was estimated that 
$200  \!<\! T_\fir \!<\! 250\,$MeV in real-world QCD, but the existence 
of finite $T_\fuv$ is uncertain at this point~\cite{Alexandru:2019gdm}. 

The above classification was inferred from the behavior of overlap spectral 
density, with overlap serving as an external rather than a dynamical probe. 
This distinction is immaterial for glue since gluonic operators can be expressed 
via the overlap matrix 
elements~\cite{Horvath:2006md, Alexandru:2008fu, Liu:2007hq} regardless 
of overlap's relation to dynamical quarks. There is however a crucial point 
to stress,  which enlightens the importance and novelty of our results.

The requirement of universality of Dirac spectra in the continuum limit
leaves only two possibilities: either the IR peak will eventually be visible 
also in the dynamical staggered operator when the continuum limit is approached,
or it will disappear also in the overlap operator in the same limit.
Our study proves that the first possibility is actually realized but it is
interesting to describe how the second could take place.
Once the dynamical-quark operator is able, close enough to the continuum limit,
to correctly resolve the low-lying structure of the spectrum,
the configurations that support an accumulation of small eigenvalues could 
be suppressed by the fermion determinant and the IR peak could disappear 
if the entropy does not overcome this suppression.
Remarkable examples of such back-reaction due to dynamical quarks
exist in the literature, e.g. related to topology, where correct continuum physics 
is recovered only at very fine lattice spacings due to large lattice 
artifacts of the staggered operator in the low-lying part of the spectrum
\cite{Bonati:2015vqz,Petreczky:2016vrs,Borsanyi:2016ksw,Frison:2016vuc,
Alexandrou:2017bzk,Burger:2018fvb,Bonati:2018blm,Lombardo:2020bvn,
Athenodorou:2022aay}.
It is therefore remarkable that staggered formulation is providing us with the first 
complete and self-consistent evidence that an IR-separated 
{\it quark-gluon medium} exists in the temperature regime of 
the proposed IR phase. This is indeed a key piece of physics conveyed by 
the present~work.

\smallskip
\noindent
{\bf 3.~Lattice Setup.}
We study $N_f \!=\! 2 +1$ QCD via lattice regularization based 
on stouted staggered Dirac operator and tree-level Symanzik-improved 
glue action. Rooting was used to control the number of quark species, 
and ensembles at physical quark masses were generated by means of 
the RHMC algorithm~\cite{Kennedy:1998cu, Clark:2006fx}. 
In certain cases, multicanonical algorithm~\cite{Berg:1992qua,Bonati:2017woi, 
Jahn:2018dke, Bonati:2018blm, Athenodorou:2022aay,Bonanno:2022dru} 
was used to enhance topological jumps without spoiling the importance 
sampling. The entire setup, including ensembles themselves, was 
adopted from Ref.~\cite{Athenodorou:2022aay}.

We focus on the thermal state at $T\!=\! 230\,$MeV which is within the region 
where the onset of IR phase was first estimated~\cite{Alexandru:2019gdm}. 
In fact, the results of recent work~\cite{Meng:2023nxf} suggest that this 
is inside the IR phase. Our results are based on ensembles described in 
Table~\ref{tab:MC_params}. The prominent role is played by those at spatial 
size $L_s \!=\! 3.43\,$fm. Note the wide range of lattice spacings in this set.

\begin{table}[!t]
\begin{center}
  \vspace{0.07in}
  \begin{tabular}{|c|c|c|c|c|c|}
  \hline
  \multicolumn{6}{|c|}{$T=230$ MeV $\simeq 1.48 \, T_c$}\\
  \hline
  &&&&&\\[-1em]
  $\beta$ & $a$~[fm] & $am_s \cdot 10^{-2}$ & $N_s^3 \times N_t$ & $L_s$~[fm] & Statistics\\
  \hline
  &&&&&\\[-1em]
  3.814 & 0.1073 & 4.27 & $32^3 \times 8$* & 3.43 & 1025 \\
  \hline
  &&&&&\\[-1em]
  3.918 & 0.0857 & 3.43 & {$40^3 \times 10$}* & 3.43 & 595 \\
  \hline
  &&&&&\\[-1em]
  4.014  & 0.0715 & 2.83 & {$48^3 \times 12$} & 3.43 & 624 \\
  \hline
  &&&&&\\[-1em]
  \multirow{3}{*}{4.100}  & \multirow{3}{*}{0.0613} & \multirow{3}{*}{2.40} & $48^3 \times 14$ & 2.94 & 224 \\
  &&& {$56^3 \times 14$}  & 3.43 & 780 \\
  &&& {$64^3 \times 14$}* & 3.92 & 228 \\
  \hline
  &&&&&\\[-1em]
  4.181  & 0.0536 & 2.10 & $64^3 \times 16$ & 3.43 & 320 \\
  \hline
  \end{tabular}
\end{center}
\vspace{-0.08in}
\caption{Ensembles of $N_f \!=\! 2\!+\!1$ lattice QCD at the physical 
point. Simulation details are given in Ref.~\cite{Athenodorou:2022aay} 
(also~\cite{Aoki:2009sc, Borsanyi:2010cj, Borsanyi:2013bia}). Bare 
parameters $\beta$ and $m_s$ are the gauge coupling and the strange 
quark mass ($m_{u,d} \!=\! m_s/28.15$). Spatial lattice size~is
$L_s \!=\! N_s a$ and $1/T \!=\! N_t a$; $a$ is the lattice spacing. 
Cases~marked by * 
did not use the multicanonical algorithm.~Gauge configurations (statistics) 
are separated by 30 RHMC trajectories.}
\vspace{-0.22in}
\label{tab:MC_params}
\end{table}

\smallskip
\noindent
{\bf 4.~Spectral Density.}
Staggered Dirac operator $D_\stag$ has imaginary eigenvalues and
the spectral equation thus takes the continuum-like form 
$D_{\stag}[U_{\stag}] u_\lambda \!=\! i \lambda u_\lambda$ with 
real $\lambda$. Here $U_\stag$ is the glue field smoothed as in 
the Monte-Carlo evolution (two steps of stout smearing). 
The eigenvalue problem was numerically solved using the PARPACK 
library, with 150 lowest positive eigenvalues obtained for each 
configuration~\cite{Athenodorou:2022aay}. Since $D_\stag$ 
quadruples the physical quark degrees of freedom, the bare 
staggered Dirac spectral density at coarse-graining width 
$\delta$ reads
\beq
  \rho(\lambda, \delta) = 
  \frac{\;\; \nu(\lambda + \delta/2) 
                \,-\,   \nu(\lambda - \delta/2) \;\;}
       {4 V \delta}
  \label{eq:020}
\eeq
where $\nu(\lambda)$ is the QCD-averaged number of modes in 
the interval $(0, \lambda]$ and $V \!=\! L_s^3/T$. Since
$\nu(\lambda)$ is RG-invariant and $\lambda$ renormalizes 
as the quark mass $m$, the renormalization of $\rho$ proceeds
via~\cite{Giusti:2008vb,Bonanno:2019xhg,Bonanno:2023ypf}
\begin{equation}
   \rho_\ren(\lambda_\ren, \delta_\ren) = 
   \frac{m}{m_\ren} 
   \rho\Bigl( \frac{m}{m_\ren} \lambda_\ren, \frac{m}{m_\ren} \delta_\ren \Bigr) 
   \label{eq:040}   
\end{equation}
The continuum limit is approached along the line of constant physics 
where $m_\ren$ remains fixed. We chose the strange quark mass 
and its current renormalized FLAG value
$m_{s\ren} \!=\! m_s^{\scriptscriptstyle{(\overline{\mathrm{MS}})}}(\mu\!=\!2\,\text{GeV}) 
\!=\! 92.2(1.0)\,\text{MeV}$ \cite{FlavourLatticeAveragingGroupFLAG:2021npn} 
to renormalize the staggered spectral density using Eq.~\eqref{eq:040}.

For 50 configurations of $N_t \!=\! 12$ system, we computed 200 lowest
eigenmodes of the overlap operator~$D_\OV[U_\stag]$ 
in the implementation used in
Refs.~\cite{Alexandru:2015fxa, Alexandru:2019gdm,Alexandru:2021pap, 
Alexandru:2021xoi} ($\rho\!=\!26/19$).

\begin{figure}[!t]
\vskip -0.00in
\centering
\includegraphics[scale=0.24]{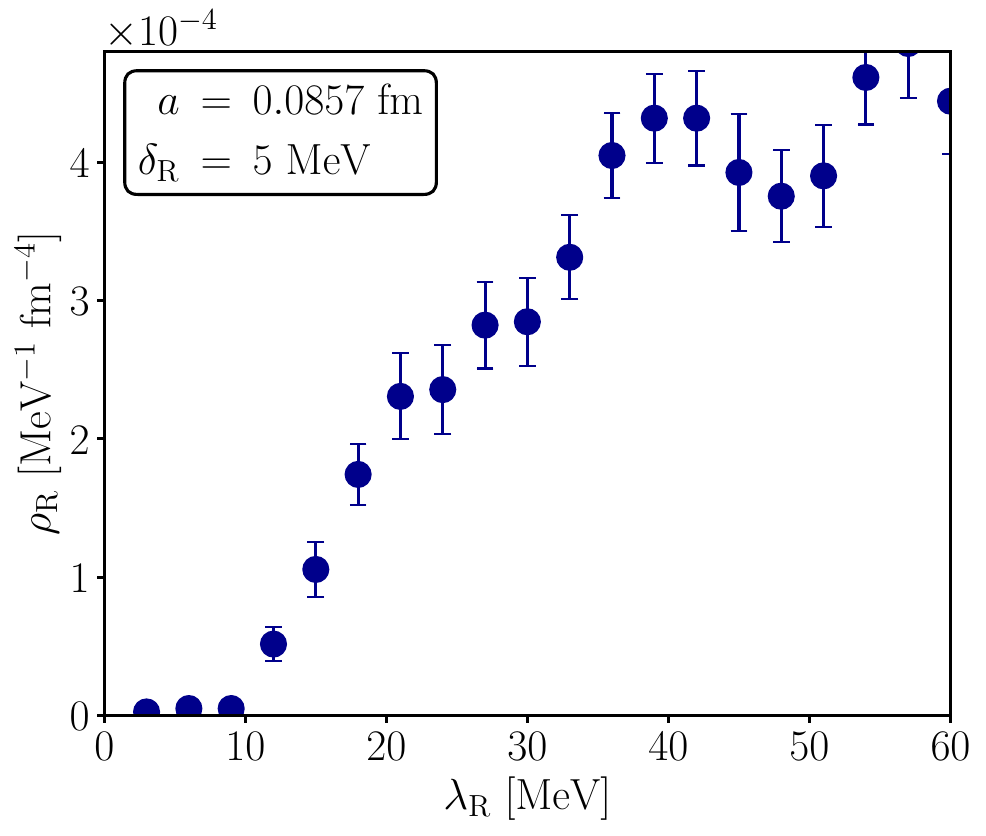}
\includegraphics[scale=0.24]{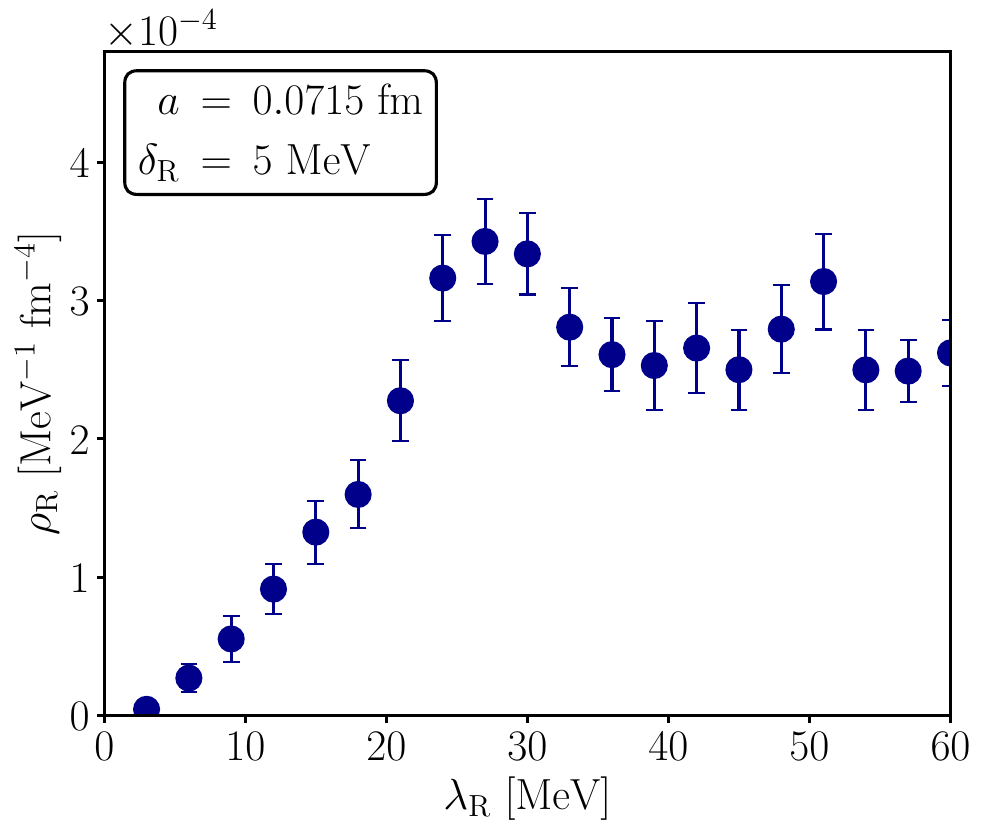}
\includegraphics[scale=0.24]{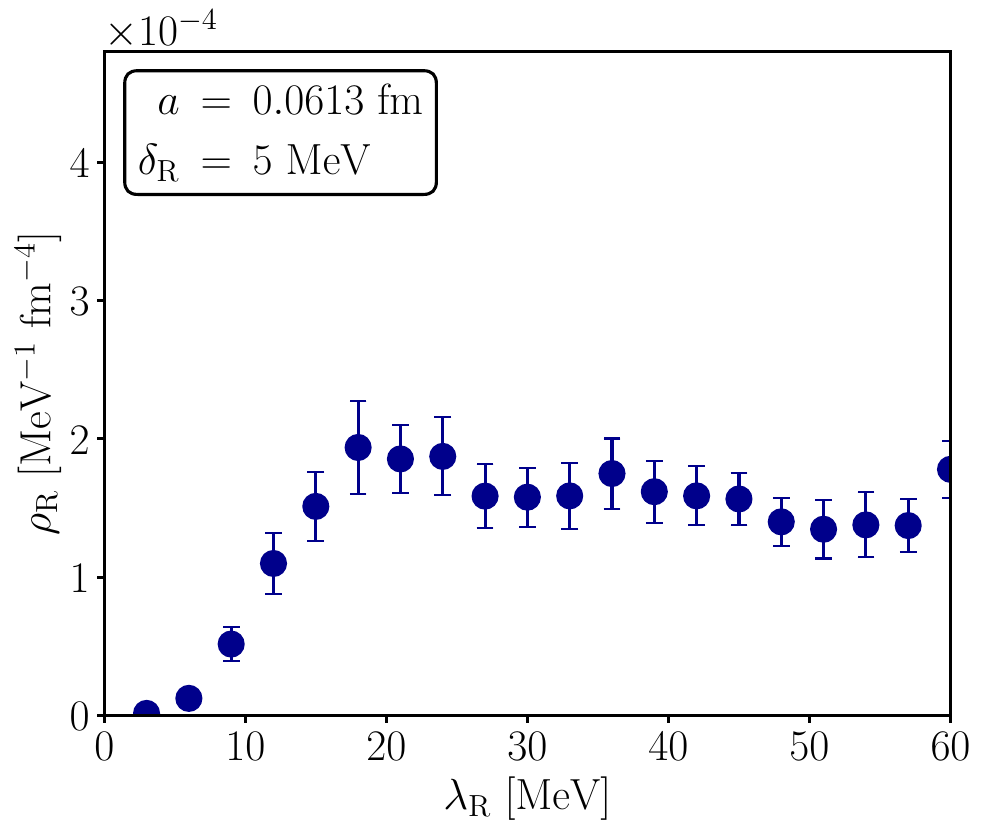}
\includegraphics[scale=0.24]{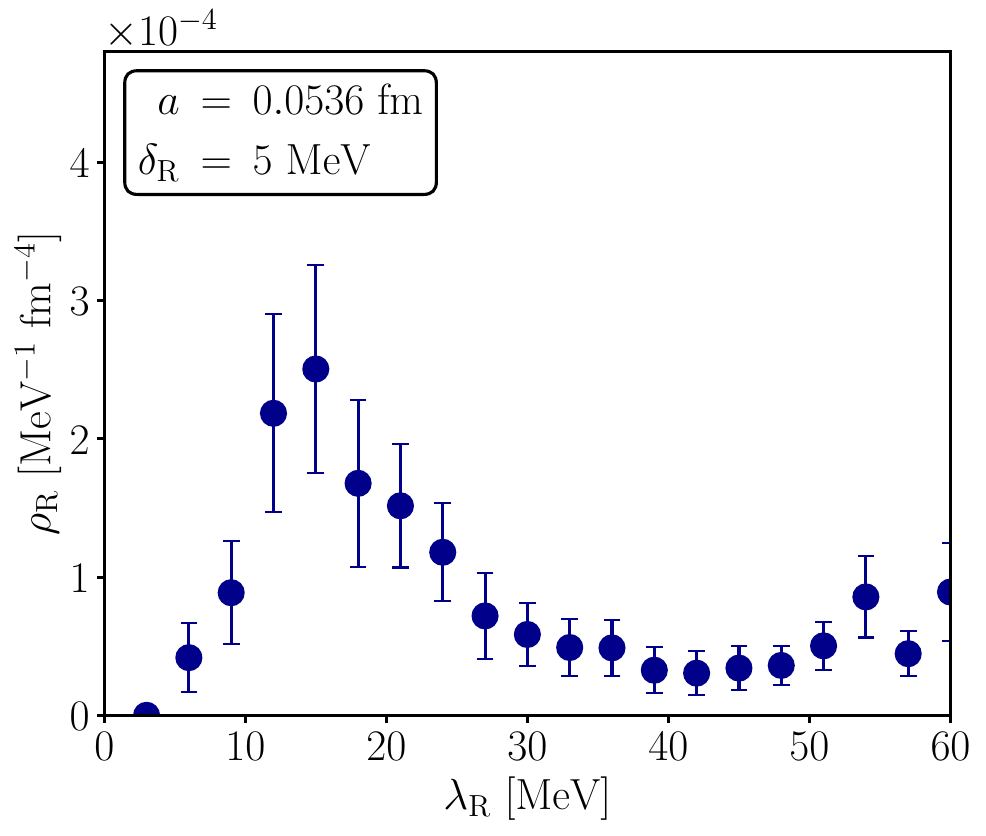}
\vskip -0.03in
\caption{Spectral densities $\rho_\ren(\lambda_\ren)$ at 
temperature $T\!=\! 230\,${\rm MeV} and spatial size 
$L_s \!=\! 3.43\,${\rm fm}. 
Coarse-graining is $\delta_\ren \!=\! 5\,${\rm MeV}.}
\vskip -0.20in
\label{fig:stag_ov_COMP}
\end{figure}

\smallskip
\noindent
{\bf 5.~Results.}
We start by showing, in Fig.~\ref{fig:stag_ov_COMP}, spectral~densities 
at decreasing lattice spacing $a$ and fixed spatial size $L_s\!=\!3.43\,$fm. 
The key observation is that the displayed low end 
of the spectrum (IR cutoff $1/L_s \!\approx\! 57\,$MeV) responds unusually 
strongly to the change of UV cutoff. While there is no sign of IR peak structure
at $a \!=\! 0.086\,$fm, the degrees of freedom redistribute as $a$ decreases, 
and the peak becomes evident at  $a \!=\! 0.054\,$fm. These trends strongly 
suggest the existence of deep-IR peak in the continuum limit.  The emerged 
structure is very similar to those previously seen in the external\ overlap operator.
Given the observed width of the peak,  the resolution 
$\delta_\ren \!=\! 5\,${\rm MeV} is appropriate to draw these conclusions,
and the stability with respect to decreasing $\delta_\ren$ was checked.

\begin{figure}[!t]
\vskip -0.00in
\centering
\includegraphics[scale=0.24]{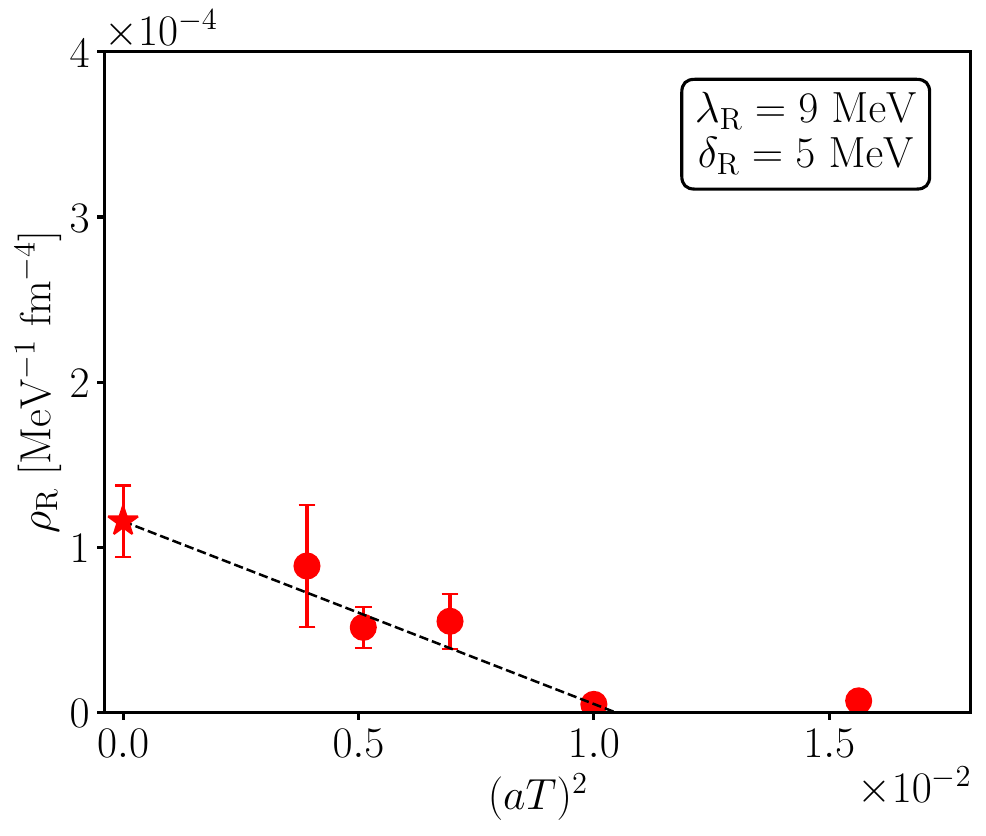}
\includegraphics[scale=0.24]{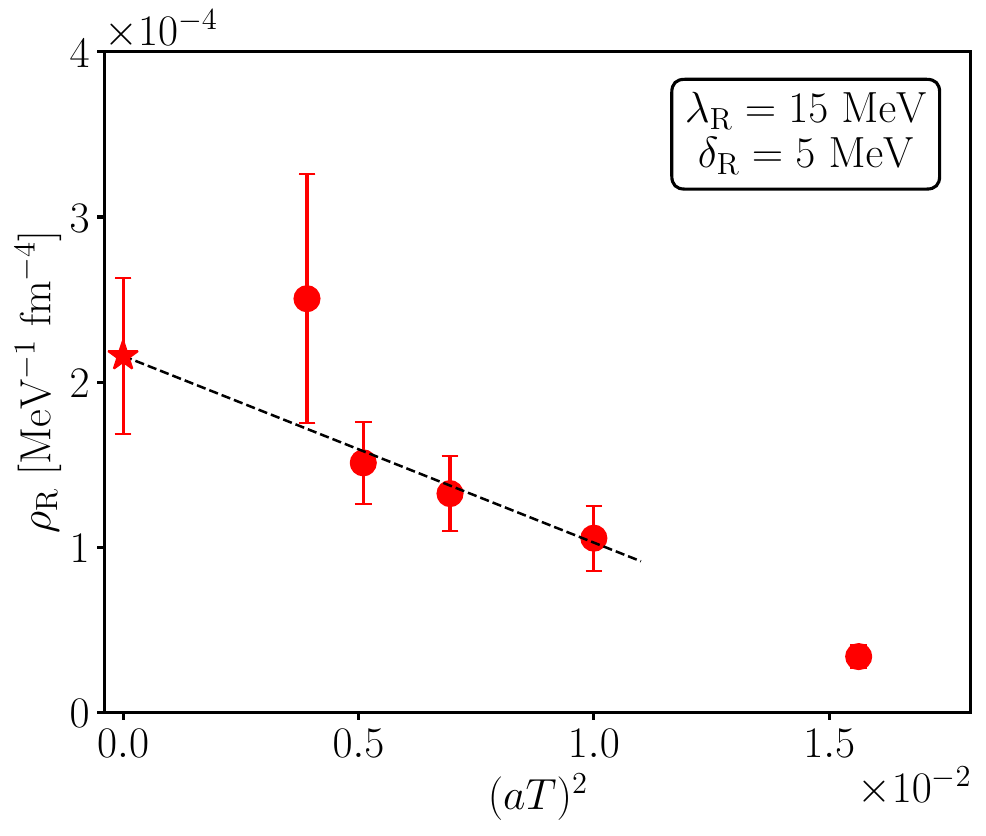}
\includegraphics[scale=0.24]{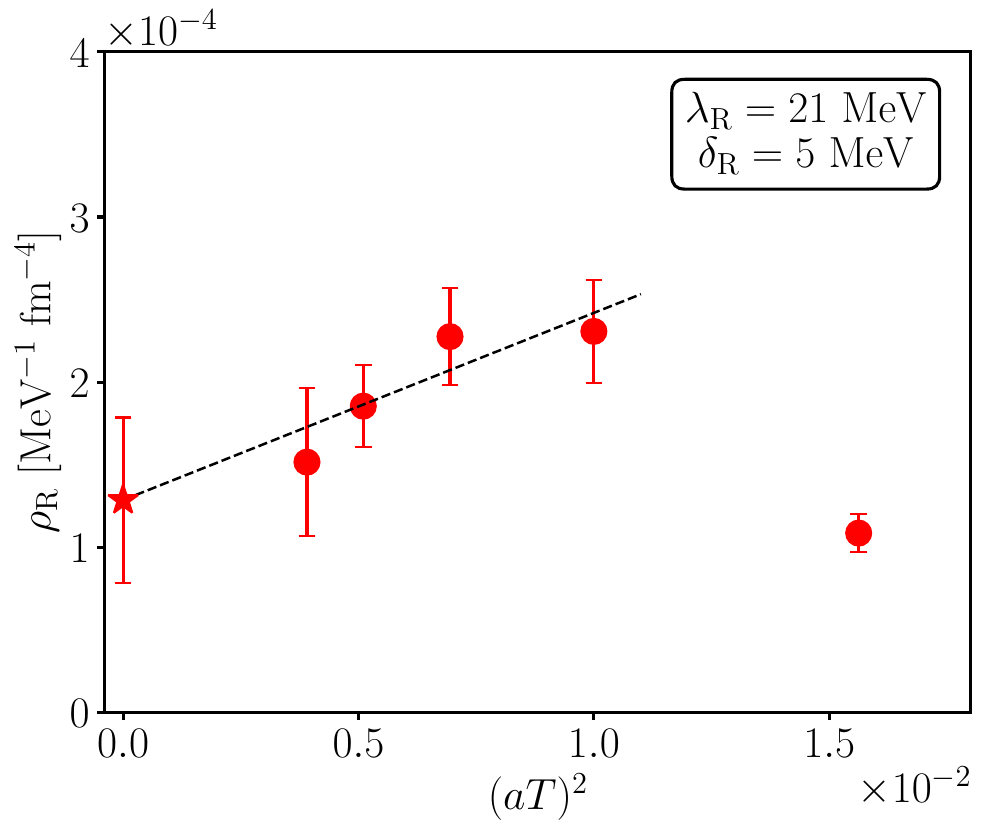}
\includegraphics[scale=0.24]{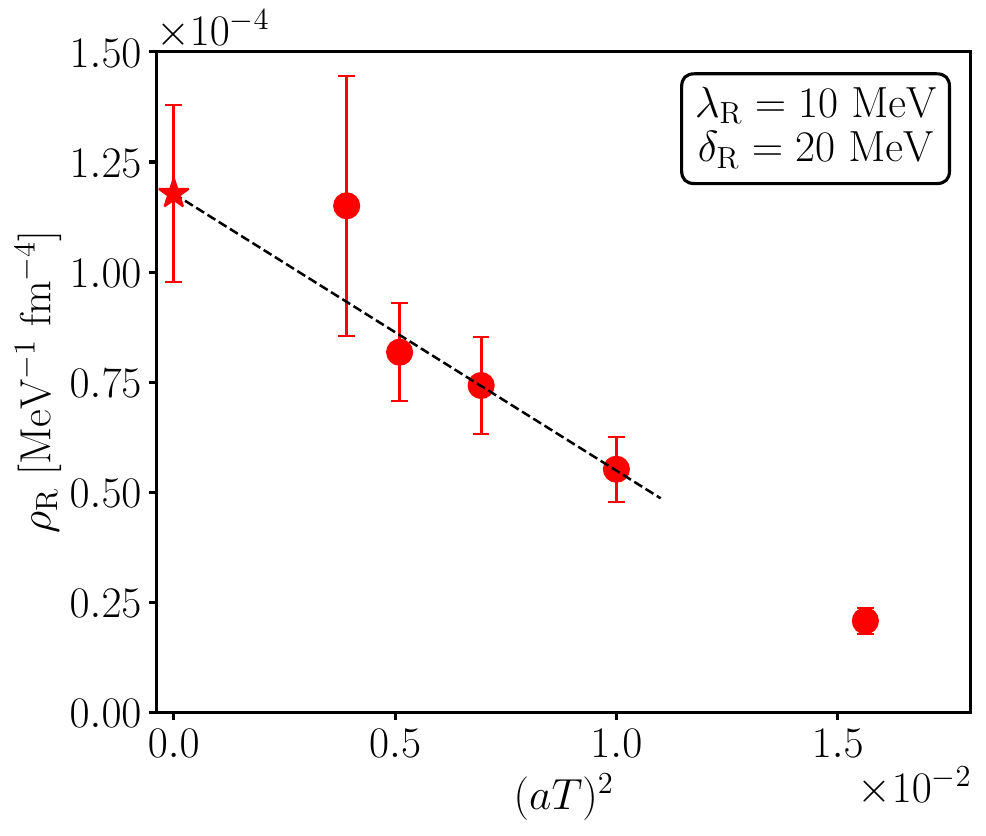}
\vskip -0.03in
\caption{Continuum extrapolations of renormalized spectral densities 
described in the text.}
\vskip -0.20in
\label{fig:cont_limit_PEAK}
\end{figure}

To corroborate this, we show in Fig.~\ref{fig:cont_limit_PEAK} 
the approach to the continuum for  $\rho_\ren(\lambda_\ren, \delta_\ren \!=\! 5\,\text{MeV})$ 
in the peak,~namely for $\lambda_\ren \!=\! 9,15,21\,$MeV. 
Extrapolations (linear in $a^2$) used the four finest lattices and 
indicate the existence of non-zero deep-IR densities in the continuum limit. 
To see this in a yet more robust way, we treat the interval 
$(0,20]\,$MeV as a single bin, i.e. evaluate $\rho_\ren$ at  
$\lambda_\ren \!=\! 10\,$MeV, $\delta_\ren \!=\! 20\,$MeV. This
is shown in the bottom-right plot of Fig.~\ref{fig:cont_limit_PEAK}.
The scale has changed here because this observable averages 
density over the entire peak region.

One could question whether a reliable continuum extrapolation for
the low-lying region of the Dirac spectrum can be obtained at all by means 
of staggered quark simulations, given the slow convergence to continuum 
for chiral observables in this particular discretization. To that end, 
we observe that reliable continuum extrapolations have been obtained 
in Refs.~\cite{Athenodorou:2022aay,Bonanno:2023thi}, with the same set 
of lattice spacings, for topological observables strongly correlated 
with the same region of the spectrum. Moreover, with the coarsest 
lattice spacing already discarded, continuum extrapolations look 
reasonably smooth and stable against removing another lattice spacing. 
With that said, we are aware of possible pitfalls related to the continuum 
extrapolation and, indeed, for the middle part of the spectrum 
(see below) we only report an upper bound of the density.

The meaning of results in Fig.~\ref{fig:cont_limit_PEAK} is analogous to those 
in Fig.~3 (bottom right) of Ref.~\cite{Alexandru:2015fxa}. Indeed, while scaling 
in Ref.~\cite{Alexandru:2015fxa} shows that decoupled IR {\it glue medium} exists 
in the continuum limit of pure-glue QCD in IR phase, Fig.~\ref{fig:cont_limit_PEAK} 
informs us that such {\it quark-glue medium} exists in the continuum limit 
of real-world~QCD at $T\!=\! 230\,$MeV.

\begin{figure}[!t]
\vskip -0.00in
\centering
\includegraphics[scale=0.24]{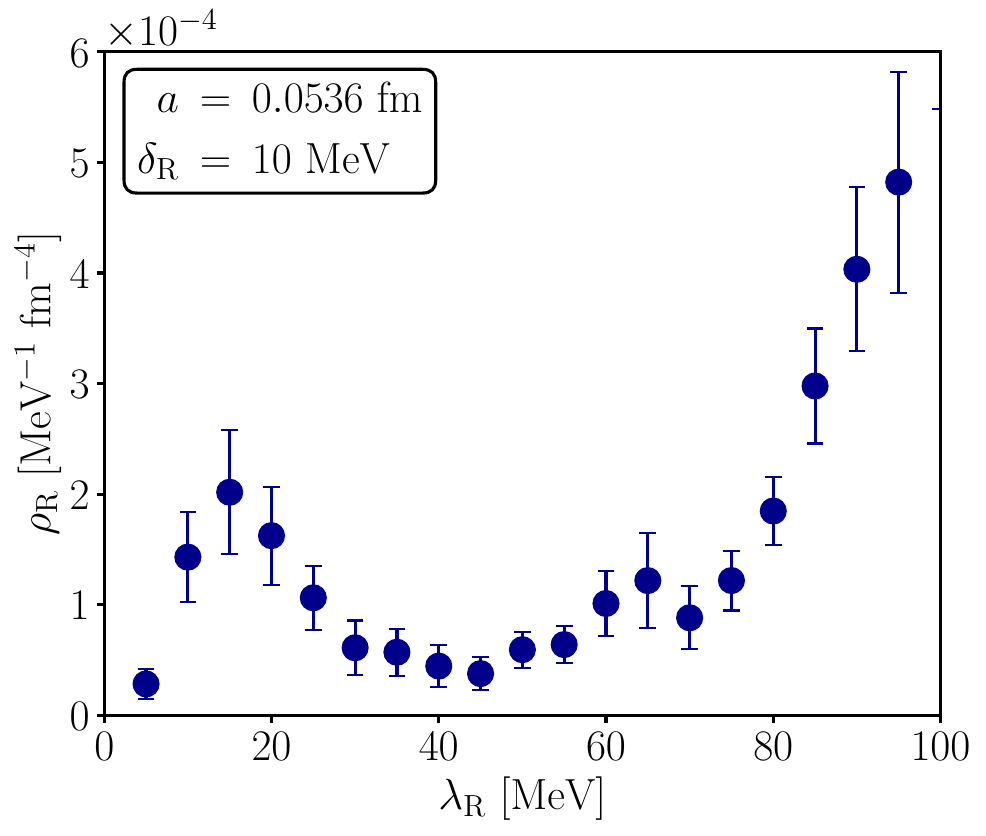}
\includegraphics[scale=0.24]{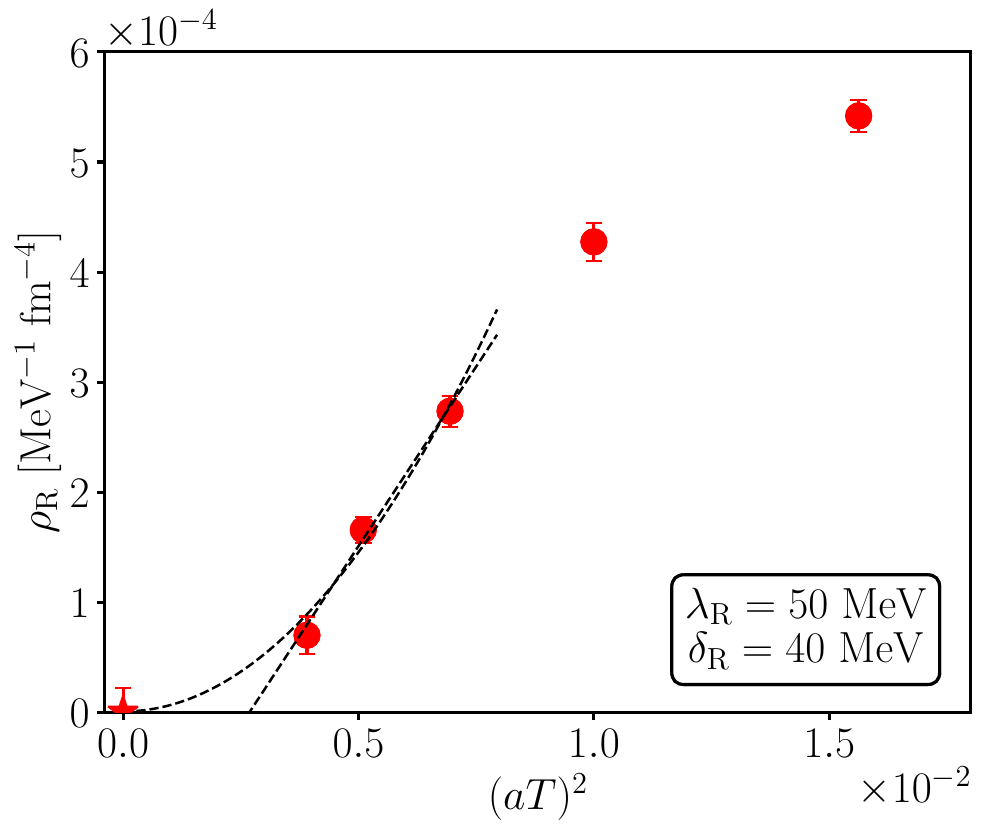}
\vskip -0.06in
\caption{Left: renormalized spectral density for $a\!=\! 0.0536\,$fm at 
$\delta_\ren \!=\! 10\,${\rm MeV}. Right: scaling of average spectral density in 
the range $\lambda_\ren \in (30,70)\,$MeV inside the plateau.}
\vskip -0.20in
\label{fig:plateau}
\end{figure}

The emergence of IR peak in the approach to the continuum involves a simultaneous 
drop of spectral density in a region to the right: a process visualized by 
Fig.~\ref{fig:stag_ov_COMP}. This creates a spectral regime of depleted density, 
referred to as ``plateau"~\cite{Alexandru:2021pap, Alexandru:2021xoi}. To identify 
its extent, we show in Fig.~\ref{fig:plateau} (left) the behavior of $\rho_\ren$ 
on a larger spectral domain and with more coarse-graining for better clarity. 
It reveals that $\lambda_\ren \in (30,70)\,$MeV is inside the plateau. 
Hence, to assess plateau's scaling properties and its degree of depletion 
in the continuum, we evaluate $\rho_\ren$ at $\lambda_\ren \!=\! 50\,$MeV 
and $\delta_\ren \!=\! 40\,$MeV. 
The scaling is shown in Fig.~\ref{fig:plateau} (right). We see that 
the decrease of density toward the continuum is so fast that the standard 
(linear in $a^2$) extrapolation would in fact lead to negative density. 
The plateau may thus mark a truly distinct regime where scaling properties 
are different (higher power). To that end, we show in Fig.~\ref{fig:plateau} 
also an extrapolation quadratic in $a^2$ which leads to value consistent 
with zero. Another possibility is that ordinary scaling hasn't settled in 
yet in the plateau. In either case, the observed behavior implies that 
the plateau is severely depleted of modes, which is consistent with
IR decoupling~\cite{Alexandru:2019gdm, Alexandru:2021pap, Alexandru:2021xoi}.

\begin{figure}[!t]
\vskip -0.00in
\centering
\includegraphics[scale=0.25]{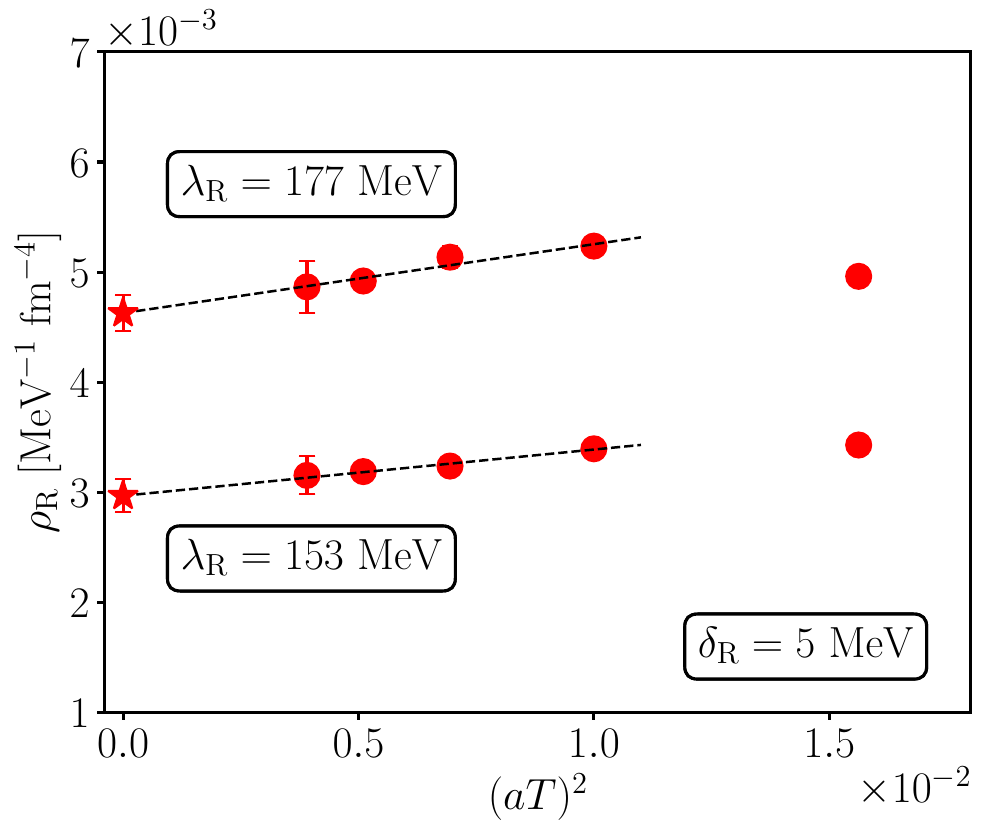}
\includegraphics[scale=0.25]{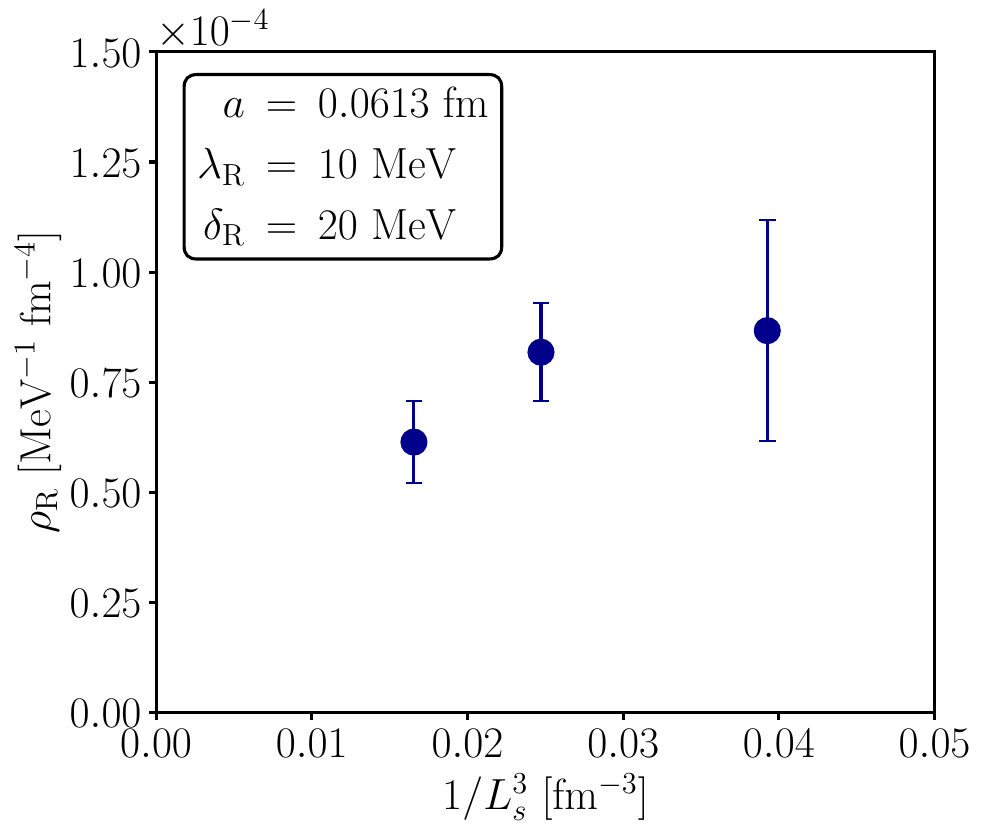}
\vskip -0.05in
\caption{Left: scaling of $\rho_\ren$ in the right-rise region
near the Anderson-like mobility edge 
$\lambda_{\mathrm{A}}\simeq 166(8)~\mathrm{MeV}$~\cite{Bonanno:2023mzj}.
Right: volume dependence of average IR-peak density at $a \!=\! 0.0613\,$fm.}
\label{fig:cont_limit_RISE}
\vskip -0.22in
\end{figure}

We now turn to the ``right rise", namely the region past the plateau, where 
density increases again. In Fig.~\ref{fig:cont_limit_RISE} we show 
the scaling at $\lambda_\ren \!=\! 153\,$MeV and $\lambda_\ren \!=\! 177\,$MeV. 
These values are near the continuum limit of the Anderson-like 
mobility edge 
$\lambda_{\mathrm{A}} \!\simeq\! 166(8)~\mathrm{MeV}$~\cite{Bonanno:2023mzj}, 
determined from the very same ensembles used here.
We observe robust (linear in $a^2$) extrapolations to finite values.

Our data allow for some assessment of volume effects. 
In Fig.~\ref{fig:cont_limit_RISE} (right) we show the average peak 
density, exhibiting a mildly decreasing trend as $L_s$ increases. 
Thus, while overlap has a strong $L_s$ and mild $a$  
dependence~\cite{Alexandru:2019gdm, Alexandru:2015fxa, 
Meng:2023nxf}, the behavior of staggered operator is the opposite. 
This not only confirms the intuitive notion of overlap ``seeing deeper 
into the continuum" due to superior chiral properties, but it also fixes 
the proper order of limits in extracting characteristics such as $p$, 
namely $\lim_{L \to \infty} \, \lim_{a \to 0} \Box$.

\smallskip
\noindent
{\bf 6.~Staggered topology and IR peak.} A particularly relevant 
question for the present work is whether the IR peak identified 
in the continuum staggered spectral density is actually due to 
an accumulation of near-zero modes. Indeed, unlike the overlap 
discretization, the staggered operator explicitly breaks chiral 
symmetry at non-zero lattice spacing and does not feature exact 
zero modes. Could it be that the observed peak effect is in fact
fully attributable to modes that can be considered ``zero modes" 
shifted to non-zero values away from the continuum limit?

While any attempts to resolve this are bound to be afflicted by 
ambiguity, certain reasonable checks can be performed. Indeed, since 
lowest staggered modes are expected to be closely related to topological 
content of the underlying gauge field, one possibility is 
to rely on heuristic assumptions rooted in the index theorem 
to approximately identify ``would-be-zero" modes. 
The continuum-extrapolated staggered spectral density can then be
recalculated with these modes removed to see whether an identifiable 
IR peak effect remains.

One such heuristic approach to identify staggered would-be-zero 
modes was given in Ref.~\cite{Borsanyi:2016ksw}. Due to taste 
degeneracy, the index theorem for the continuum limit of the theory 
with staggered quarks reads $4 Q=n_{\rm left} - n_{\rm right}$, 
relating topological charge $Q$ and numbers of left- and right-handed 
zero modes. Assuming this is realized by the minimum number 
of zero modes, the spectrum of $iD_{\rm stag}[U]$ should possess 
$2 \vert Q[U] \vert$ positive would-be-zero modes 
(the other $2 \vert Q[U] \vert$ are negative due to the staggered 
symmetries) sufficiently close to the continuum limit. The prescription 
of~\cite{Borsanyi:2016ksw} is to identify them as the first 
$2\vert Q[U] \vert$ positive lowest-lying modes.

In order to determine the topological content of a gauge configuration,
in this study we relied on the gluonic calculation of the topological charge 
of Ref.~\cite{Athenodorou:2022aay}, performed on the same set of 
configurations, using the so-called $\alpha$-rounded clover definition 
computed after cooling: 
$Q=\mathrm{round}\left[\alpha \, Q_{\rm clov}^{(\mathrm{cool})}\right]$, 
with $\alpha$ chosen so as to minimize the averaged squared distance of 
the cooled clover lattice charge $Q_{\rm clov}^{(\mathrm{cool})}$ from 
integers. The reader is referred to the dedicated 
study~\cite{Athenodorou:2022aay} for further details.

\begin{figure}[!t]
\vskip -0.00in
\centering
\includegraphics[scale=0.235]{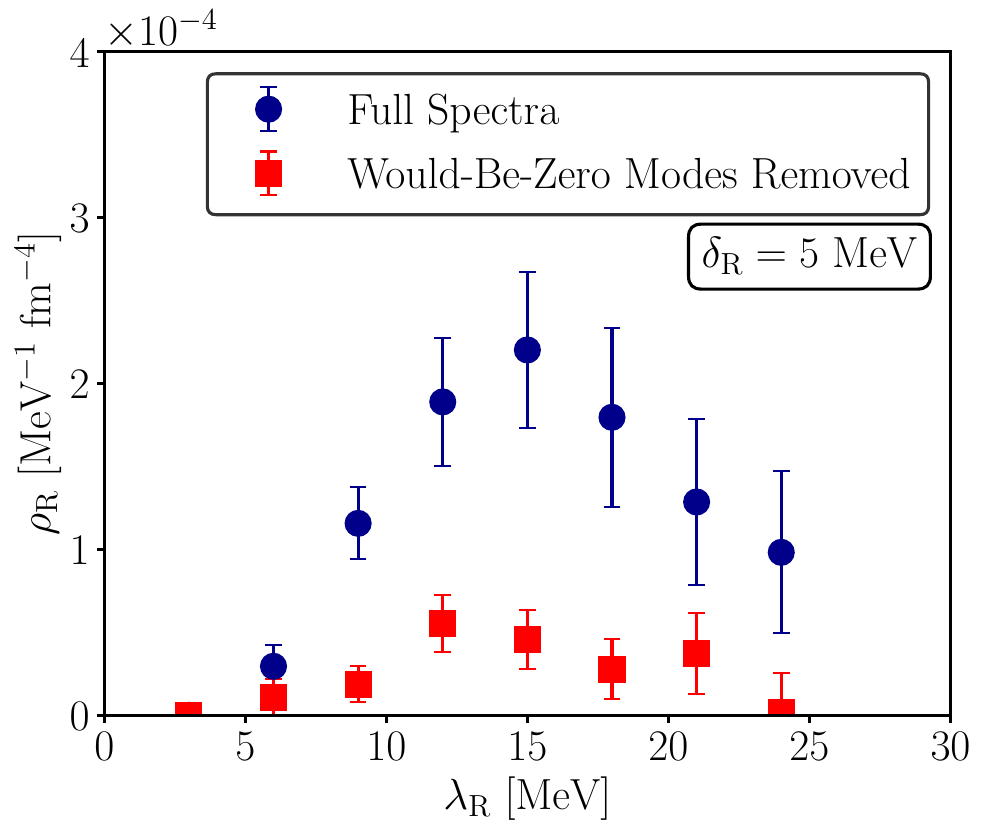}
\includegraphics[scale=0.235]{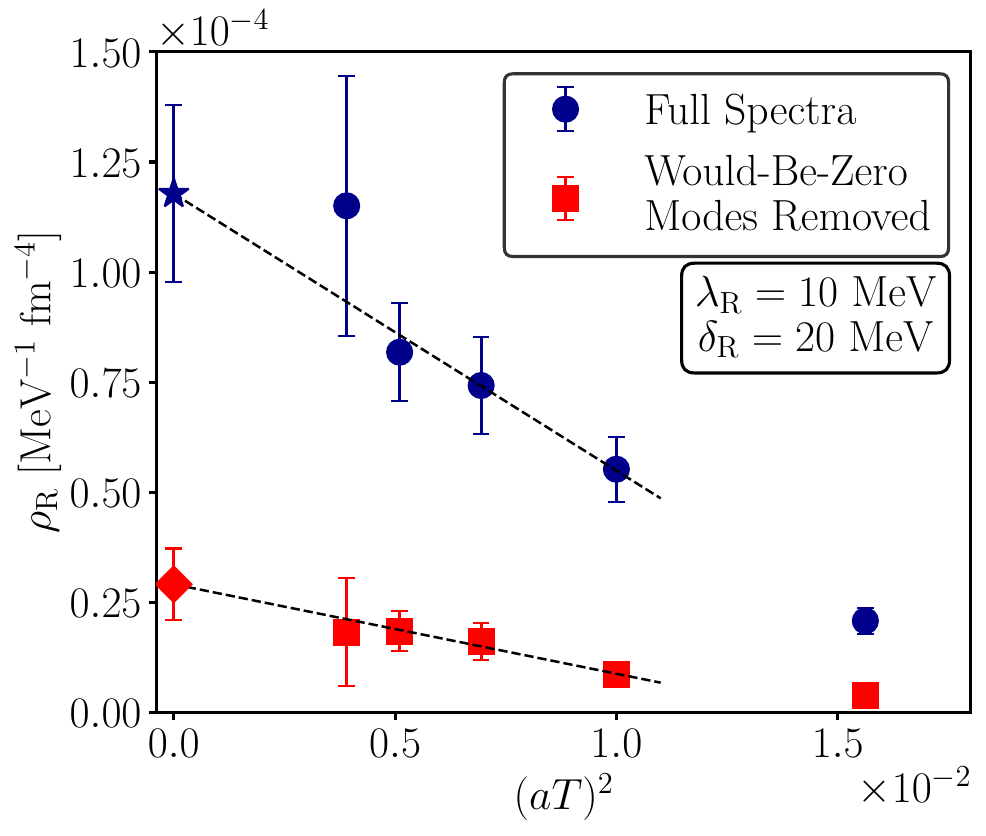}
\includegraphics[scale=0.235]{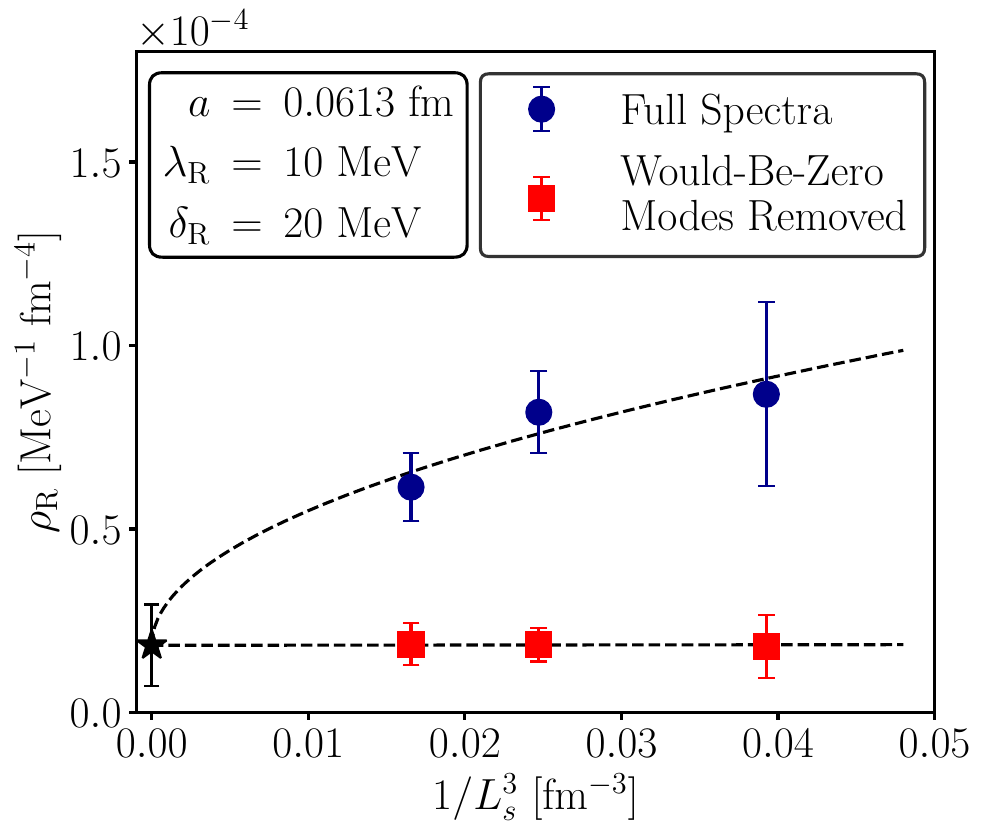}
\vskip -0.05in
\caption{Top left: comparison of the continuum-extrapolated 
spectral densities in the peak region before and after the removal of 
possible topological would-be-zero modes (see text) from staggered 
spectra. 
Top right: continuum extrapolation of the spectral density in 
a large bin $\lambda_{\ren}\in[0,20]$ MeV including the peak region 
before and after the removal of possible topological would-be-zero 
modes from staggered spectra. Bottom: peak-averaged spectral density 
for $N_t=14$ (corresponding to $a=0.0613$ fm) as a function of 
the inverse spatial volume before and after the removal of 
possible topological would-be-zero modes from staggered spectra.}
\label{fig:wbzms_removed}
\vskip -0.22in
\end{figure}

The results obtained after the identification and subtraction of 
would-be-zero modes from staggered spectra are shown in 
Fig.~\ref{fig:wbzms_removed}. In the top left plot, we compare 
the continuum-extrapolated results for $\rho_{\ren}$ obtained in 
the range $\lambda_\ren \in [0,25]$ MeV with a bin size $\delta_{\ren}=5$ MeV. 
These data show that a good fraction of the peak could indeed be 
attributed to would-be-zero modes. However, the signal does not vanish 
in the continuum limit even after the removal of such modes and we 
conclude that our full data indeed reflects a non-trivial accumulation 
of low-lying near-zero modes. Similar conclusions can be drawn from 
the continuum limit of the spectral density calculated at 
$\lambda_{\ren}=10$ MeV and $\delta_{\ren} = 20$ MeV 
(Fig.~\ref{fig:wbzms_removed}, top right), corresponding to one large 
bin covering most of the peak region. Indeed, we observe a clear 
non-vanishing signal for $\rho_{\ren}$ in the continuum limit,
both before and after the removal of possible would-be-zero modes.

The fact that the value of $\rho_\ren$ at the peak after the removal 
of would-be-zero modes is reduced is consistent with the slightly 
descending trend observed in the peak-averaged density as a function 
of the inverse spatial volume $1/L_s^3$ shown in 
Fig.~\ref{fig:cont_limit_RISE} on the right. As a matter of fact, 
the contribution of would-be-zero modes to the spectral density is 
expected, at fixed temperature, to be suppressed in the thermodynamic 
limit as $1/L_s^{3/2}$, due to the index theorem, meaning that it is less 
and less important on larger volumes.

This observation leads to a possible interesting counter-check, 
in order to strengthen our analysis: if the residual peak really accounts 
for the presence of a physical IR structure, related to near-zero modes, 
it should become dominant in the thermodynamic limit, since 
the would-be-zero modes contribution vanishes in that limit. Unfortunately 
we have not performed, for computational reasons, extensive simulations 
at different spatial volumes for all values of the lattice spacing.

However we can consider the case reported in Fig.~\ref{fig:cont_limit_RISE} 
on the right, corresponding to $a = 0.0613$~fm, 
for which we have three different volumes. After repeating the removal 
of would-be-zero modes in all cases, we obtain the results for 
the peak height as a function of $L_s$ shown in bottom panel of 
Fig.~\ref{fig:wbzms_removed}, where they are directly compared to those 
before the removal: the residual peak appears to be practically independent 
of the volume, or rather slightly rising as a function of $1/L_{\rm s}^3$, 
while the original peak decreases as $1/L_{\rm s}^{3/2}$, so that the two 
data sets suggest a nice convergence to the same thermodynamic 
limit,~as~expected.

As a final comment, we stress that we verified that compatible results 
are obtained, and thus similar conclusions can be drawn, if, instead 
of removing the lowest-lying modes according to the prescription 
of~\cite{Borsanyi:2016ksw}, the spectral density is projected onto 
the $Q=0$ topological sector, a procedure which is expected to remove 
the contribution of would-be-zero modes too.

\begin{figure}[!t]
\vskip -0.00in
\centering
\includegraphics[scale=0.235]{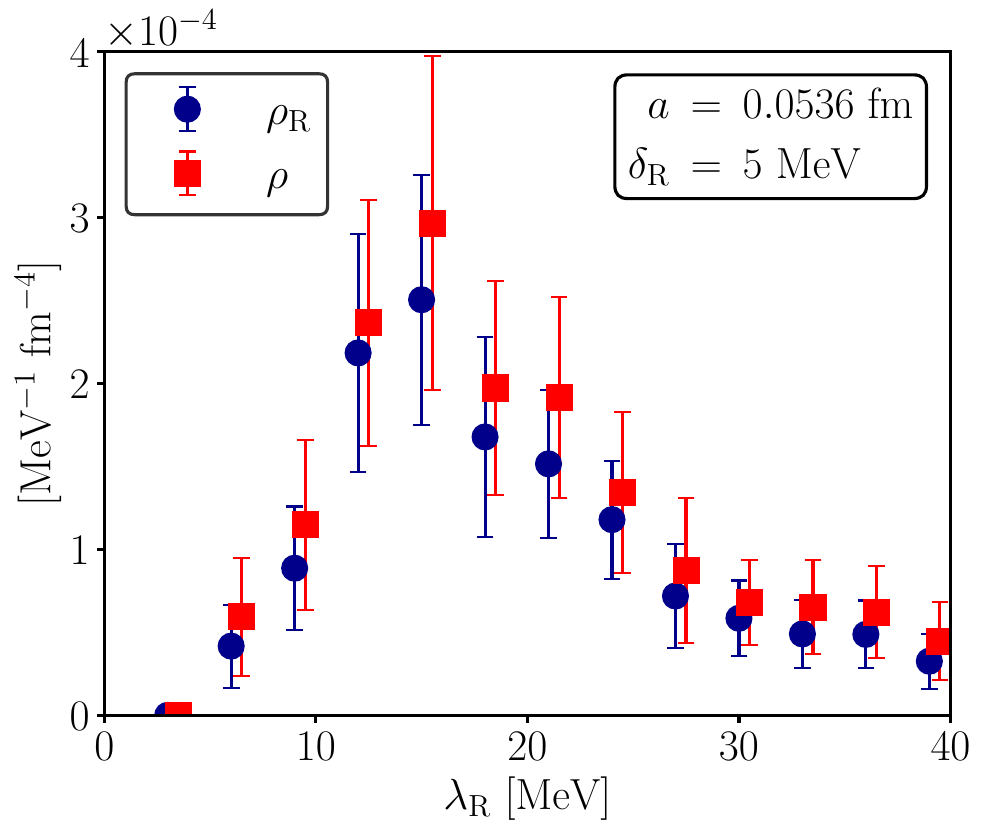}
\includegraphics[scale=0.235]{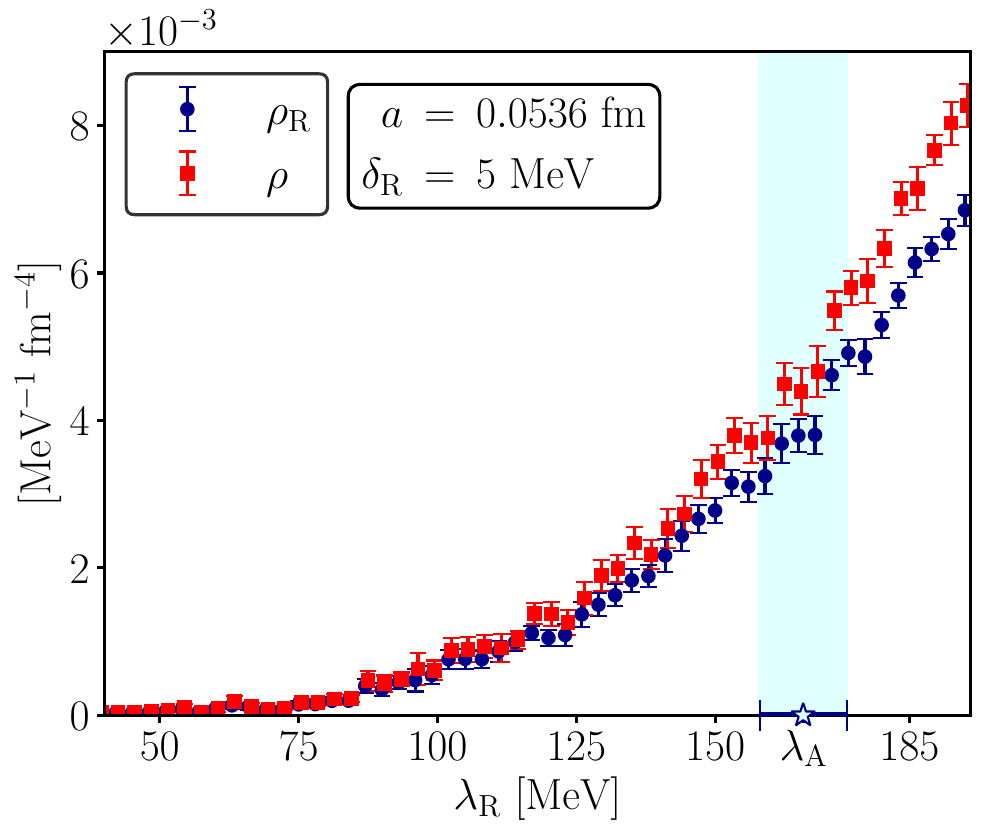}
\caption{Bare and renormalized spectral densities in the left rise (left) 
and the complement (right) on the finest lattice.}
\vskip -0.1in
\label{fig:bare_vs_renorm_COMP}
\end{figure}

\smallskip
\noindent
{\bf 7.~Renormalization effects.} It was suggested~\cite{Alexandru:2019gdm} 
that renormalization won't affect the IR part of the IR-bulk separated system 
due to the IR-bulk decoupling. This was later associated with Anderson-like 
criticality~\cite{Alexandru:2021pap, Alexandru:2021xoi} inducing non-analyticities 
at the separation scale(s) and preventing bulk fluctuations from influencing 
the IR. How is this expressed formally via Eq.~\eqref{eq:040}? 
Strict non-renormalization would require $\rho(\lambda) \!=\! c/\lambda$ in 
the left rise, since then $\rho(\lambda_\ren) \!=\! \rho_\ren(\lambda_\ren)$. 
A negative power $p \!=\! -1 +\delta$, $\delta$ 
small~\cite{Alexandru:2019gdm}, only leads to weak rescaling.  
To test this, in Fig.~\ref{fig:bare_vs_renorm_COMP} we compare $\rho$ and 
$\rho_\ren$ in the left rise (left) and the rest (right). 
The two dependencies differ at large scales but come together roughly below 
the Anderson-like point $\lambda_{\rm A}$~\cite{Bonanno:2023mzj}, 
and become statistically indistinguishable in the left rise and the plateau.

\smallskip
\noindent {\bf 8.~Discussion and Conclusions. $\,$} We computed 
the renormalized Dirac spectral density in $N_f \!=\!2+1$ QCD at physical 
quark masses and temperature $T \!=\! 230\,$MeV, on a system of size 
$L_s \!\simeq\!3.4~$fm. To that end, we extrapolated the staggered 
density of lattice QCD with staggered quarks to the continuum limit, 
using cutoffs down to $a\!\simeq\! 0.054\,$fm. 

Our work is unique both in terms of the temperature regime explored 
and the results obtained. Regarding the former, $T \!=\! 230\,$MeV 
is well above the chiral crossover at 
$T_c \!\simeq\! 155\,$MeV~\cite{Aoki:2006we,Aoki:2006br,Aoki:2009sc,
Borsanyi:2010bp,Bhattacharya:2014ara,HotQCD:2018pds,Borsanyi:2020fev} 
and near the center 
of predicted window $200 \!<\! T_\fir \!<\! 250\,$MeV for the onset 
of IR phase~\cite{Alexandru:2019gdm}. The point-wise continuum limit 
of $\rho_\ren(\lambda_\ren)$ has not been computed before in 
the high-temperature regime $T \!>\! T_c$ of real-world QCD, 
although certain continuum characteristics near $T_c$ were extracted 
in Ref.~\cite{Kaczmarek:2023bxb}.
This has required probing very fine lattice spacings, in view of the slow 
approach to continuum of the dynamical staggered operator, especially 
for properties regarding the low lying part of the spectrum.

Our key new result is the numerical proof that the IR peak structure 
like one previously seen in the overlap spectral density of pure-glue 
and real-world QCD with staggered and Wilson quarks
(see \cite{Edwards:1999zm}, \cite{Alexandru:2015fxa, Dick:2015twa}, 
\cite{Alexandru:2019gdm,Alexandru:2021pap, Alexandru:2021xoi},
\cite{Meng:2023nxf} for milestones), also exists in 
the {\it dynamical-quark} density of real-world QCD. 
%
We also exhibited clear numerical evidence that, after removing the contribution of 
would-be-zero modes, this peak structure survives the continuum limit as well, 
meaning that near-zero modes give a non-vanishing contribution to it. Moreover, 
we collected evidence, even if only at one lattice spacing where we had simulations 
for three different spatial volumes available, that the residual peak becomes 
dominant in the thermodynamic limit, meaning that it can indeed be associated with 
a physical IR structure associated with near-zero modes, whose contribution 
is expected to survive the thermodynamical limit, contrary to what happens 
for would-be-zero modes.

\begin{figure}[!t]
\centering
\includegraphics[scale=0.355]{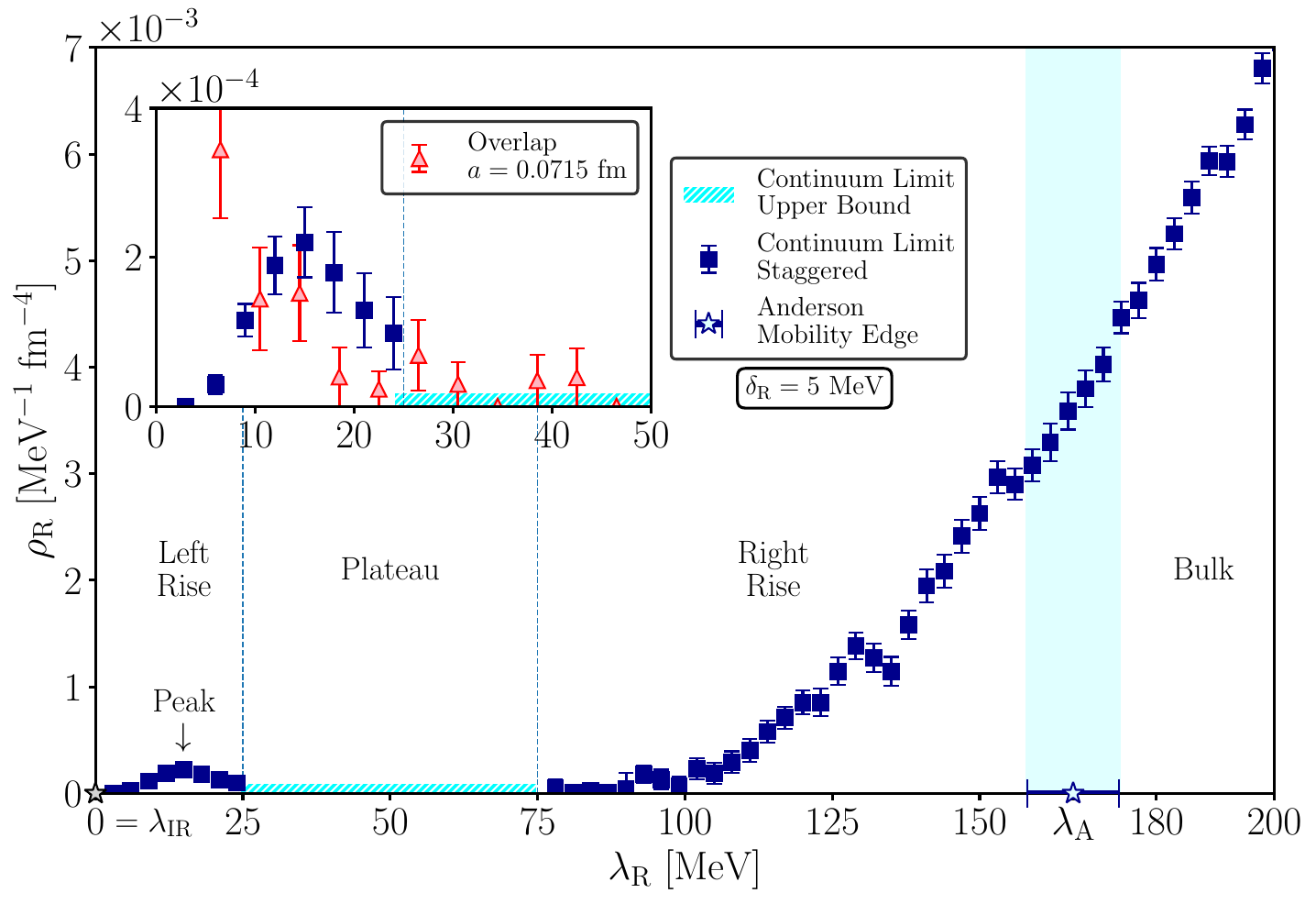}
\caption{Continuum limit for $\rho_{\ren}(\lambda_{\ren})$ in 
$N_f \!=\! 2+1$ QCD at $T \!=\! 230\,${\rm MeV} and 
$L_s \!=\! 3.4\,${\rm fm}. Anderson-like mobility edge 
$\lambda_{\mathrm{A}}\simeq 166(8)\,${\rm MeV} was obtained 
from the same ensembles in Ref.~\cite{Bonanno:2023mzj}. 
$\lambda_\fir \!=\!0$ is the IR mobility edge predicted
in Ref.~\cite{Alexandru:2021xoi}.}
\vskip -0.25in
\label{fig:cont_limit_rho}
\end{figure}

Our results have significant consequences in two physical 
contexts~\cite{Alexandru:2015fxa, Dick:2015twa}.
The first one~\cite{Alexandru:2015fxa} developed 
the idea~\cite{Alexandru:2012sd, Alexandru:2014paa} to classify 
phases of gauge theories with fundamental quarks without invoking
chiral symmetry breaking. The precise form of this was given in 
Ref.~\cite{Alexandru:2019gdm} where the observation of strong 
negative near-pure power in well-separated overlap IR peak 
($\rho(\lambda) \!\propto\! \lambda^p$ for $\lambda \!\to\! 0$ 
with $p \!\approx\!-\! 1$) led to the definition of IR phase as a regime 
where $p <0$, and to its association with decoupled IR glue and
elements of IR scale invariance. The IR structure found
here in the {\it dynamical-quark} staggered operator is fully analogous.
Indeed, the continuum limit, summarized by Fig.~\ref{fig:cont_limit_rho}, 
exhibits signature features of the left rise (peak), the plateau and 
the right rise~\cite{Alexandru:2015fxa, Alexandru:2021pap, 
Alexandru:2021xoi,Meng:2023nxf}, with severe depletion of modes in 
the plateau. Fits near the top of the left rise 
give $p \!\approx\! -1.3(2)$ ($a\!=\! 0.0536\,$fm) and, although the present 
setup/data does not allow for a more robust estimate, the strong 
separation of the peak, its renormalization properties, and the overall 
similarity to previous overlap results suggest that the system is in IR 
phase and that physical quarks couple to the separated IR glue. 
Thus, QCD at $T\!=\!230\,$MeV features an IR component made up
of both glue and quark fields: a {\it quark-gluon} thermal~medium.

The second aspect connects the IR peak to the problem of thermal 
U$_\mathrm{A}$(1) anomaly restoration~\cite{Dick:2015twa}. 
Although this concerns the chiral limit of QCD, results in the light-quark 
regime, such as ours, are relevant to the problem at this stage.
Indeed, the current debate in this area deals with the very
issue outlined above: is there an IR peak in the dynamical-quark
spectral density at light quark masses~\cite{Aoki:2020noz, 
Ding:2020xlj, Kaczmarek:2023bxb, Kaczmarek:2021ser, Aoki:2021qws}?  
Our results show that the IR structure emerges near 
the continuum limit which puts such doubts to rest. Moreover, it 
is known empirically~\cite{Alexandru:2014zna, Alexandru:2014paa, 
Alexandru:2015fxa, Alexandru:2019gdm} that the effect of lowering 
the quark mass is to strengthen the IR peak. We thus do not expect 
substantial weakening of the anomalous effect toward the chiral limit.
In that sense, our results are consistent with the anomalous nature
of U$_\mathrm{A}(1)$ at $T\!<\!230\,$MeV, as found also in 
Ref.~\cite{Ding:2020xlj} (see also~\cite{Kovacs:2023vzi}). Nevertheless, 
only simulations explicitly examining trends toward the chiral limit can 
ultimately decide this question.

A few more points are well-worth making.

\noindent 
{\it (i)} Our results fully validate, by properly taking into account 
the quark back-reaction, the observation of IR phase ($p\!<\!0$) 
previously obtained via the external overlap operator, which is 
capable of detecting IR features also on coarser lattices.
The inset of Fig.~\ref{fig:cont_limit_rho} shows the overlap 
density for comparison, featuring an IR peak at UV cutoff 
where staggered IR structure hasn't developed yet 
($a \!=\! 0.0613\,$fm) on only 50 gauge configurations. 
Renormalization for this case is described in the Appendix.

\noindent 
{\it (ii)} Our results suggest that the predicted transition 
temperature $T_\fir$ is below 230 MeV (see also~\cite{Meng:2023nxf}). 
Combination with the original estimate~\cite{Alexandru:2019gdm} 
gives $200 \!<\! T_\fir \!<\! 230\,$MeV. The implication here is that 
the IR peak in the range $T_c \!<\! T \!<\! T_\fir$, which is known 
to exist at $T \!=\! 175\,$ MeV~\cite{Alexandru:2015fxa}, 
is only logarithmic ($p \!=\!0$) like at low temperatures. 
Dynamical overlap calculations in this temperature 
range~\cite{Kotov:2024xpn} should bring an additional insight into 
this issue.

\noindent 
{\it (iii)} The stochastic model of Ref.~\cite{Kovacs:2023vzi} extends 
that of Ref.~\cite{Edwards:1999zm}. It treats determinant effects via 
reweighting (within the model) from pure glue. It appears to naturally 
generate $p\!<\!0$, and may thus offer a useful tool to model the IR 
features akin to those studied here.    

\smallskip
\noindent {\bf Acknowledgments.$\,$} A.~A.~is supported in part by the U.S. 
DOE Grant No.~DE-FG02-95ER40907. The work of C.~B.~is supported by the Spanish 
Research Agency (Agencia Estatal de Investigación) through the grant IFT 
Centro de Excelencia Severo Ochoa CEX2020-001007-S and, partially, by grant 
PID2021-127526NB-I00, both funded by MCIN/AEI/10.13039/501100011033. 
C.~B.~also acknowledges support from the project H2020-MSCAITN-2018-813942 
(EuroPLEx) and the EU Horizon 2020 research and innovation programme, 
STRONG-2020 project, under grant agreement No.~824093. C.~B.~acknowledges 
useful discussions with Matteo Giordano. This work has also been partially supported 
by the project ”Non-perturbative aspects of fundamental interactions, in the Standard 
Model and beyond” funded by MUR, Progetti di Ricerca di Rilevante Interesse 
Nazionale (PRIN), Bando 2022, grant 2022TJFCYB (CUP I53D23001440006). 
Numerical calculations have been 
performed on the \texttt{Marconi} and \texttt{Marconi100} machines at Cineca, 
based on the agreement between INFN and Cineca, under projects INF22\_npqcd and 
INF23\_npqcd.

\bibliographystyle{apsrev4-2}
\bibliography{refs}

\appendix
\section*{Appendix: Overlap-Staggered Matching}

In order to properly renormalize the overlap spectral density, we performed 
matching between staggered and overlap quarks. We followed the strategy 
proposed in Ref.~\cite{Kaczmarek:2021ser}, which looks for overlap masses 
$m_s^{(\OV)}$ and $m_{u,d}^{(\OV)}=m_s^{(\OV)}/28.15$ such that the quantity
\beq
\Delta = \frac{ m_s^{(\OV)} \langle\bar{\psi}\psi\rangle_l - m_{u,d}^{(\OV)} 
\langle\bar{\psi}\psi\rangle_s }{T^4}
\eeq
where $\langle\bar{\psi}\psi\rangle_f \equiv 
\langle\bar{\psi}\psi\rangle(m^{(\OV)}_f)$, is equal between the two 
fermion discretizations.

For staggered quarks, chiral condensate was computed by means of noisy 
estimators. For the overlap, instead, we computed the quantity 
$a^3 \langle\bar{\psi}\psi\rangle(m)$ following the same prescription 
as in Ref.~\cite{Kaczmarek:2021ser}, namely:
\beq
\left\langle \sum_{\lambda_\OV\ne0} 
\frac{\frac{2(am)}{N_s^3N_t}\left[4(aM)^2-\frac{\vert a \lambda_\OV\vert^2}{(aM)^2}\right]}
{\frac{\vert a \lambda_\OV \vert^2}{(aM)^2}[4(aM)^2-(am)^2] + 4 (am)^2 
(aM)^2}\right\rangle
\eeq
where $aM\! \!\equiv\! \rho \!=\!26/19$ is the negative mass kernel parameter of 
$D_\OV$,  $\lambda_\OV$ are the bare overlap eigenvalues, and the sum was carried 
over up to the first $\mathcal{O}(100)$ eigenvalues.

Following this method, we find that $a m_s^{(\OV)} \simeq 0.0286$ for 
the $48^3 \times 12$ lattice, to be compared with the corresponding 
staggered strange quark mass $a m_s = 0.0283$.
The two masses thus turn out to be in a remarkable agreement.

We finally mention here that the magnitude of a complex overlap eigenvalue 
$\lambda_\OV$ was used as a measure corresponding to staggered 
$\lambda$ at low~energy.

\end{document}